\documentclass[aps, twocolumn,floatfix,pre,showpacs, superscriptaddress]{revtex4-2}

\usepackage{amssymb,amsmath,amsfonts}
\usepackage{graphicx,graphics}
\usepackage{epsfig}
\usepackage{epstopdf}
\usepackage{amsfonts}
\usepackage{bm,bbm}
\usepackage{amssymb,amsmath,amsfonts}
\usepackage{mathrsfs}
\usepackage{calc}
\usepackage{epsfig}
\usepackage{float}
\usepackage{xcolor}
\usepackage{epstopdf}
\usepackage{bm,amssymb,amsmath}
\usepackage{graphicx,color}

\usepackage{ulem}
\usepackage{xspace}

\usepackage[colorlinks=true, citecolor=teal, urlcolor=purple,linkcolor=purple]{hyperref}

\begin{document}

\title{Fluctuation-response relation for a nonequilibrium system\\ with resolved Markovian embedding}

\author{R\'emi Goerlich}
\email{remigoerlich@tauex.tau.ac.il}
\affiliation{Raymond \& Beverly Sackler School of Chemistry, Tel Aviv University, Tel Aviv 6997801, Israel}

\author{Antoine Tartar}
\affiliation{University of Strasbourg and CNRS, CESQ and ISIS, UMR 7006, F-67000 Strasbourg, France}

\author{Yael Roichman}
\email{roichman@tauex.tau.ac.il}
\affiliation{Raymond \& Beverly Sackler School of Chemistry, Tel Aviv University, Tel Aviv 6997801, Israel}
\affiliation{Raymond \& Beverly Sackler School of Physics and Astronomy, Tel Aviv University, Tel Aviv 6997801, Israel}

\author{Igor M Sokolov}
\email{igor.sokolov@physik.hu-berlin.de}
\affiliation{Institut f\"ur Physik, Humboldt Universi\"at zu Berlin, Newtonstrasse 15, 12489 Berlin, Germany}

\date{\today}

\begin{abstract}
 Fluctuation–response relations must be modified to describe nonequilibrium systems with non-Markovian dynamics. Here, we experimentally demonstrate that such relation is quantitatively recovered when the appropriate Markovian embedding of the dynamics is explicitly resolved. Using a colloidal particle optically trapped in a harmonic potential and driven out of equilibrium by a controlled colored noise, we study the response to a perturbation of the stiffness of the confining potential. While the reduced dynamics violates equilibrium fluctuation–response relations, we show that the dynamical response to the stiffness perturbation is fully determined by steady-state correlations involving the exact conjugate observable in the Markovian embedding. 
 \end{abstract}


\maketitle

\textit{Introduction}\---\
For a system in thermal equilibrium, Fluctuation–Response Relations (FRRs) provide a direct connection between the spontaneous fluctuations of an observable and its response to an external perturbation \cite{takahasi1952generalized, kubo1966fluctuation, marconi2008fluctuation, caprini2021fluctuation}.
In the canonical example, the autocorrelation function of the position $x(t)$ of a diffusing object determines the dissipative mean response $\langle x(t) \rangle$ following the switching off of a constant force \cite{lenk1967simple}.
Many natural and artificial systems, however, experience perturbations of a different nature, extending beyond the constant-force paradigm.
Cells, for instance, are frequently exposed to dynamic mechanical changes in their environment \cite{guvendiren2012stiffening} or in their internal mechanical properties \cite{hurst2021intracellular}.
A sudden stiffening of the substrate constitutes a perturbation not in the form of an applied external force, but rather through a modification of the mechanical properties of the environment, which triggers a dynamical response, such as increased cell spreading.
More broadly, cellular adaptation through the recovery of homeostasis exemplifies complex perturbations that modify the system’s properties and connect states characterized by different free energies and timescales \cite{agozzino2020cells}.
Such responses, which combine nonequilibrium dynamics and complex perturbations, can be challenging to predict.

\begin{figure}[t!]
	\centerline{\includegraphics[width=1\linewidth]{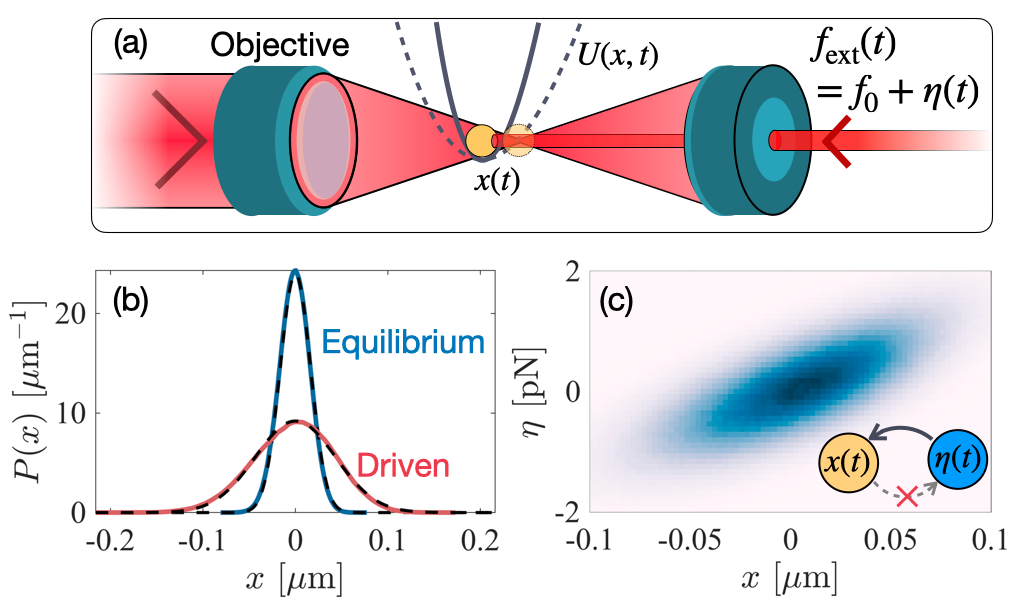}}
	\caption{(a) Schematic view of the experimental setup: a micron-sized particle suspended in water is optically trapped by a focused laser beam. It is subjected to a time-dependent external forcing $f_{\rm ext}(t)$, imposed by the radiation pressure of a secondary laser beam. The position $x(t)$ along the optical axis is recorded via the diffraction of an third low-power laser beam (not represented).
    Further details are provided in Appendix \ref{App:Exp} and in Refs. \cite{goerlich2022harvesting, Goerlich2025FT}.
    (b) Experimentally measured positional probability density functions $P(x)$ in thermal equilibrium (blue) and in the presence of colored noise driving (red), together with the respective analytical Gaussian distributions (black dashed line)
    (c) Experimentally measured bivariate distribution $p(x, \eta)$ maintained out of equilibrium by the non-reciprocal coupling between both variables, as sketched in inset.}
	\label{fig:System}
\end{figure}

This motivates the search for explicit forms of FRRs applicable to broader classes of perturbations.
A recent step in this direction is the experimental verification of an FRR adapted to a perturbation of the stiffness of a harmonic potential \cite{Goerlich2025FT}.
While this FRR correctly predicts the response to a dynamical compression, it breaks down when the system is driven into nonequilibrium states by colored noise, highlighting the limits of equilibrium-inspired FRRs.
A second step toward predicting the response of complex biophysical processes is therefore to derive FRRs for systems driven out of equilibrium and subject to active fluctuations \cite{turlier2016equilibrium, maggi2014generalized, goerlich2022harvesting}.
Since the timescales at play in driven nonequilibrium systems are usually not well separated, it is generally not possible to rely on an effective temperature with clearly defined regimes \cite{puglisi2017temperature}.
When possible, recovering FRRs for nonequilibrium systems nevertheless allows one to predict the evolution of a system following a perturbation, based on the analysis of its spontaneous fluctuations in steady state \cite{prost2009generalized, dinis2012fluctuation}.
These generalized FRRs, together with the progressive inclusion of complex perturbations, lead to improved control of systems across scales, ranging from biophysics \cite{dinis2012fluctuation} to granular materials \cite{gnoli2014nonequilibrium} and climate dynamics \cite{lucarini2023theoretical}, where predicting the response to perturbations is critical.

Here, we employ a generalization of FRR that holds for a system driven out of equilibrium by a correlated noise \cite{prost2009generalized} and focus on perturbation in the form of a change in stiffness \textit{i.e.} the mechanical properties of the confining potential, which defines the system's timescale and free-energy.
The relation between fluctuations and response relies on the identification of the appropriate conjugate variable, which takes here a nontrivial form.
We show this relation experimentally by trapping a colloidal particle in a time-dependent optical potential under the action of correlated, noisy external forces.
In most biological experiments, nonequilibrium is induced by the action of a hidden degree of freedom, which is usually not accessible to the experimentalist \cite{dinis2012fluctuation}.
This challenge traces back to the state estimation problem, already present in Kalman filtering \cite{kalman1961new, bechhoefer2015hidden}.
In contrast, here we measure simultaneously the position of the particle and the external noise and this allows us to probe experimentally the Markovian embedding.
This model experiment reproduces the dynamics of minimalist active matter systems, while offering the possibility of precise control of the potential stiffness.
Our results show that, upon monitoring the correct conjugate variable, it is possible to predict the response of an active system to a mechanical transformation.
We also provide a formal derivation of the generalized FRR, solely based on the hypothesis of Markovianity on the system.\\

\textit{Response to a perturbation of the stiffness} \---\
In our experimental setup (see Fig.~\ref{fig:System}(a)), a single colloidal particle is trapped in a harmonic optical trap with stiffness $\kappa(t)$, which is varied in time by controlling the trapping laser's intensity \cite{goerlich2022harvesting, pires2023optimal, Goerlich2025FT}.
A secondary laser beam imposes an external radiation pressure force $f(t) = f_0 + \eta(t)$, sum of a constant contribution $f_0$, and a zero-mean exponentially correlated noise $\eta(t)$, with correlation function $\langle \eta(t) \eta(s) \rangle = \sigma_\eta^2 e^{-|t-s|/\tau_c}$ with correlation time $\tau_c$.
The contribution of the constant force $f_0$ vanishes by subtracting the mean of the recorded trajectories.
The overall dynamics of the system can therefore be described as a non-reciprocally (unidirectionally) coupled pair of equations, for the variables $x$ and $\eta$ that reads
\begin{equation}
    \begin{aligned}
        \dot x(t) &= -\frac{\kappa(t)}{\gamma} x(t) + \sqrt{2D}\xi(t) + \frac{1}{\gamma} \eta(t)\\
        \dot \eta(t) &= -\frac{1}{\tau_c} \eta(t) + \sqrt{2 D_{\eta}}\theta(t)
    \end{aligned}
    \label{Eq:Langevin}
\end{equation}
where $\gamma$ is Stokes drag coefficient, $D=k_{\rm B}T/\gamma$ is the thermal diffusion coefficient with $k_{\rm B}$ being the Boltzmann constant and $D_{\eta} = \sigma_{\eta}^2/\tau_c$ is the diffusion coefficient of the noise's dynamic.
$\xi(t)$ and $\theta(t)$ are uncorrelated zero-mean white noises.
The stiffness of the potential and the viscous drag define the system's characteristic relaxation time $\tau_r = \gamma / \kappa$.
The non-reciprocal coupling between the variables $x$ and $\eta$ drives the system in a nonequilibrium steady-state (NESS) \cite{loos2020irreversibility}.
Yet the Gaussian nature of the driving noise $\eta(t)$ maintains the Gaussian statistics $P(x)$ for $x(t)$ inthe driven system, simply characterized by a larger variance than in the absence of driving (Fig.~\ref{fig:System}~(b)) which can be captured as an effective temperature \cite{Goerlich2025FT}.

The reduced dynamics $x(t)$ is non-Markovian due to the memory contained in the colored noise.
It also vividly reveals its nonequilibrium nature by breaking the FDR of the second kind \cite{kubo1966fluctuation}: the viscosity is memory-less while the noise is correlated, as observed for bacterial baths \cite{maggi2017memory}.
Yet the possibility to record the time-series of position and noise simultaneously allows us to study the two-dimensional embedding $\{x(t), \eta(t)\}$, which is a Markovian nonequilibrium system \cite{sandford2017pressure}.
The bivariate distribution $p(x,\eta)$ is a Gaussian NESS with non-vanishing correlations between both variables (Fig.~\ref{fig:System}(c)) and stationary currents.

\begin{figure}[t!]
	\centerline{\includegraphics[width=1\linewidth]{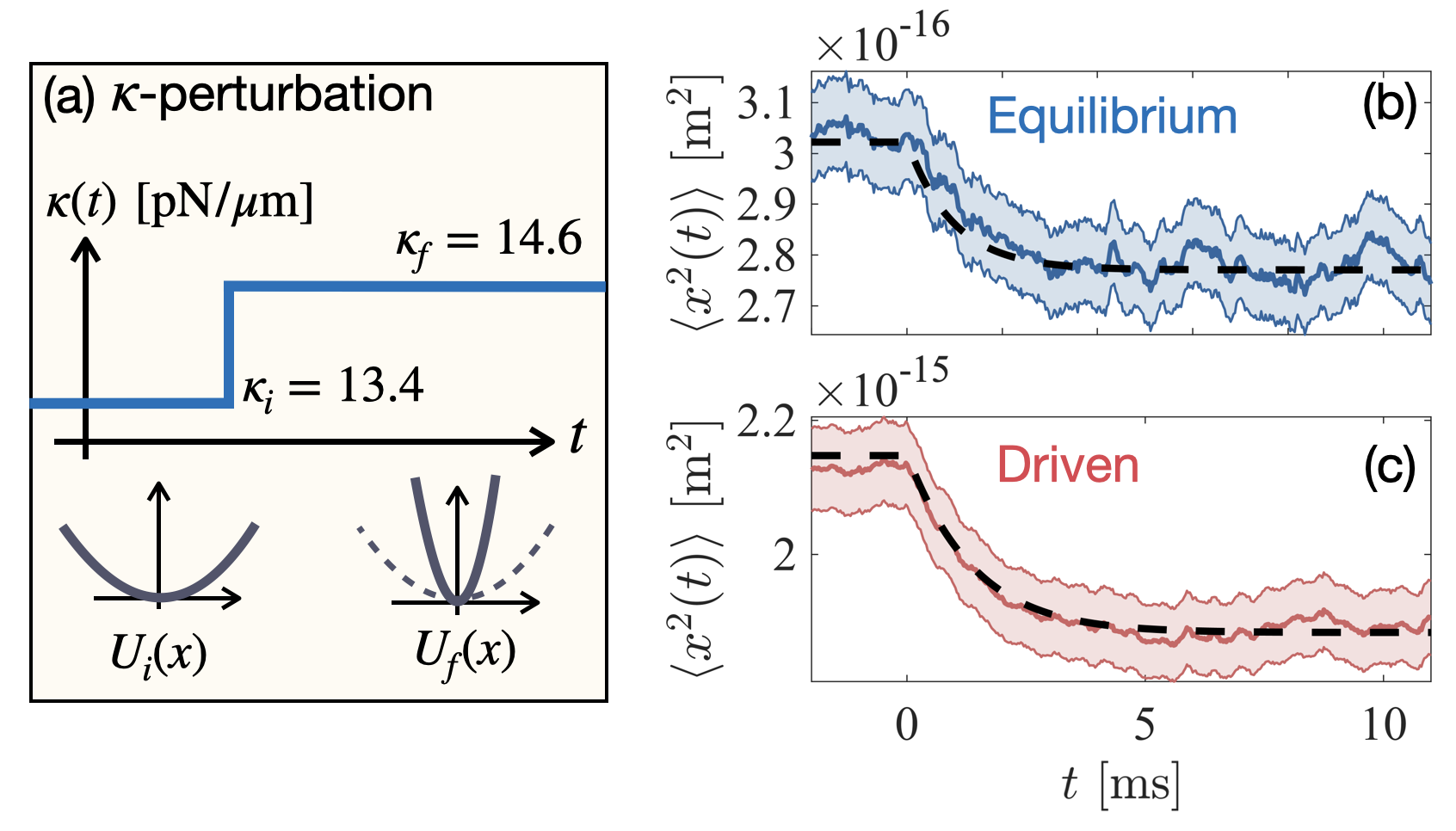}}
	\caption{(a) Schematic representation of the perturbation applied: a sudden change of the stiffness $\kappa$ of the optical potential $U(x) = \kappa x^2/2$ trapping the particle.
    (c) Time-dependent second moment $\langle x^2(t) \rangle$ following a sudden increase in $\kappa$ in the absence of driving noise, together with the analytical simple exponential decay (black dashed line).
    (c) Time-dependent second moment in the presence of exponentially correlated noise with constant variance and correlation time, together with the analytical bi-exponential decay (black dashed line).}
	\label{fig:Perturbation}
\end{figure}

By controlling the intensity of the trapping beam, we impose an abrupt change in stiffness, from $\kappa_{\rm i} = 13.4 \pm 0.11$ to $\kappa_{\rm f} = 14.6 \pm 0.13 ~\rm{pN/\mu m}$ (Fig.~\ref{fig:Perturbation}(a)) of small amplitude.
This compression (followed by its time-reversed expansion from $\kappa_{\rm f}$ to $\kappa_{\rm i}$) is applied sequentially to the trapped microparticle, with a waiting time between each repetition larger than all timescales $\tau_{\rm i} = \gamma/\kappa_{\rm i}$, $\tau_{\rm f} = \gamma/\kappa_{\rm f}$ and $\tau_c$.
We record the time-series $x(t)$ of positions of the particle subjected to this series of protocols and split it into sub-trajectories centered around each protocol, relying on the ergodicity of the sub-trajectories \cite{goerlich2021noise}.
This leads to an ensemble of \textit{circa} $18.000$ independent trajectories experiencing the same compression.
Our experimental setup, therefore, enables the collection of large statistics similar to numerical simulations, with high temporal resolution.
Both are generally challenging in biological experiments.
The response of the ensemble $\{x_j(t)\}$ to the compression $\kappa(t)$ is determined by its positional probability density $P(x,t)$.
Importantly, since at all times $P(x,t)$ is a centered Gaussian distribution, it is fully characterized by its second moment $\sigma^2(t) = \langle x^2(t)\rangle$.

In Fig.~\ref{fig:Perturbation}(b), we show the evolution of $\sigma^2(t)$ for a stiffness perturbation in the absence of driving noise $\eta(t)$.
At the beginning of the sequence, the system is at thermal equilibrium and its second moment obeys the equipartition theorem in the harmonic potential $\sigma_{\rm i}^2 \equiv \langle x^2(t<0)\rangle = k_{\rm B}T/\kappa_{\rm i}$.
At the end of sequence, the system reaches equilibrium again, with $\sigma_{\rm f}^2 \equiv \langle x^2(t\gg\tau_{\rm f})\rangle = k_{\rm B}T/\kappa_{\rm f}$.
Its dynamical evolution for finite times agrees well with the simple exponential decay predicted by solving the Fokker-Planck equation in a harmonic potential $\sigma^2(t) = \sigma_{\rm i}^2 + (\sigma_{\rm f}^2 - \sigma_{\rm i}^2)e^{-2t/\tau_{\rm f}}$ \cite{risken1989fokker}.
In Fig.~\ref{fig:Perturbation}(c), we show the response to the same stiffness perturbation with a colored noise driving $\eta(t)$ with $\tau_c = 3~\rm{ms}$.
In that case as well, the system starts and ends in a Gaussian steady state, but the respective second moment does not obey the equipartition theorem \cite{maggi2014generalized, goerlich2022harvesting}.
The evolution of $\langle x^2(t) \rangle$, under the influence of both timescales $\tau_{\rm f}$ and $\tau_c$, differs from a simple exponential decay.
It agrees well with the analytical solution for a colored-noise driven Ornstein-Uhlenbeck process (given in Appendix \ref{App:Var}), with mixed timescales $\tau_{\rm i}$, $\tau_{\rm f}$, and $\tau_c$.
In the following, we ask if these time evolutions after a $\kappa$-perturbation can be predicted from steady-state measurements.\\

\begin{figure}[t!]
	\centerline{\includegraphics[width=1\linewidth]{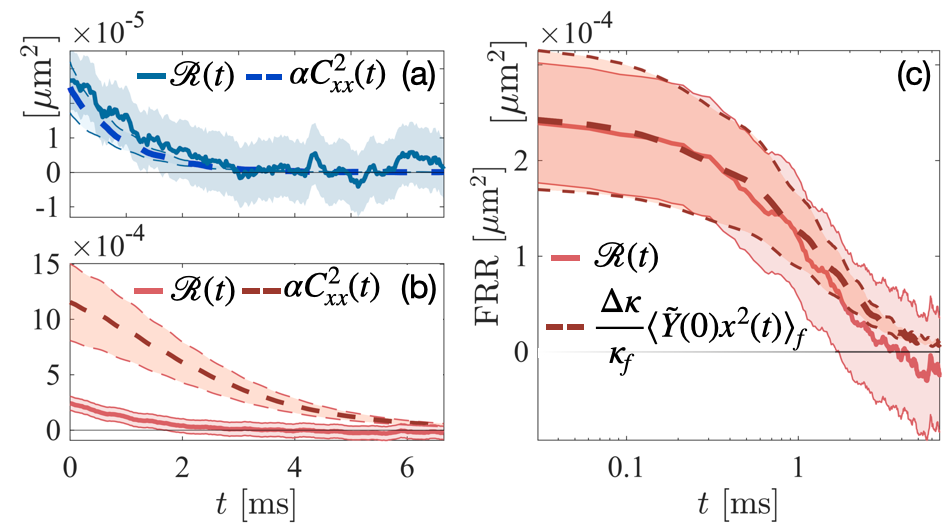}}
	\caption{(a) FRR relation Eq.(\ref{Eq:EqFRR}) for a system in thermal equilibrium, probed via a stiffness-perturbation. The response function $\mathscr{R}(t)$  (blue line) agrees with the scaled squared correlation function (dark blue dashed line).
    (b) The same FRR breaks in the presence of colored-noise driving as shown in Ref.~\cite{Goerlich2025FT}.
    (c) Generalized FRR Eq.(\ref{Eq:GFRR}) taking into account both position and noise variable: the response function $\mathscr{R}(t)$ (red line) agrees with the cross-correlation between the observable $x^2$ and its exact conjugate variable $Y$, multiplied by $\kappa_{\rm f}$ (dark red dashed line).}
	\label{fig:FRR}
\end{figure}

\textit{Experimental verification of the generalized FRR}\---\
At thermal equilibrium, the system's response to an external perturbation is directly related to its equilibrium correlation function via the temperature $T$.
Here, a stiffness perturbation leads to a response of the observable $x^2$ and the appropriate FRR should connect $\mathscr{R}(t) \equiv \sigma^2(t) - \sigma_{\rm f}^2$ to the correlation function $\langle x^2(t)x^2(0)\rangle_{\rm f}$, where $\langle ...\rangle_{\rm f}$ denotes an average in the final steady-state.
Using Wick's theorem, the latter can be expressed as a function of the square of the positional correlation function $C_{xx}(t) = \langle x(0) x(t) \rangle_{\rm f}$, leading to the following FRR for a stiffness perturbation
\begin{equation}
    \mathscr{R}(t) = \frac{\Delta \kappa}{k_{\rm B}T} \left( 1 + \frac{\Delta \kappa}{\kappa_{\rm i}}\right) C_{xx}^2(t)
    \label{Eq:EqFRR}
\end{equation}
with a first and second order contributions in the perturbation amplitude $\Delta \kappa = \kappa_{\rm f} - \kappa_{\rm i}$ \cite{Goerlich2025FT}.
In Fig.~\ref{fig:FRR}(a), we show that this FRR is fulfilled for a stiffness change between two equilibrium states: the response and correlation function share the same time-dependence and their relative amplitude is dictated by the prefactor $\alpha = \frac{\Delta \kappa}{k_{\rm B}T} \left( 1 + \frac{\Delta \kappa}{\kappa_{\rm i}}\right)$.
In Fig.~\ref{fig:FRR}(b), we, however, show that, in the presence of colored noise, the FRR is broken.
The response and correlation have distinct amplitudes, but most importantly, different time dependencies, as was already demonstrated in Ref.~\cite{Goerlich2025FT} for this case.
As a consequence of this, no constant effective temperature can allow to recover the FRR but instead, a time-dependent effective temperature is necessary to enforce the equality of $\mathscr{R}(t)$ and $\alpha C_{xx}^2(t)$ at all times \cite{dieterich2015single} and the thermodynamic meaning of such temperature in a system with mixed timescales is not clear \cite{gnoli2014nonequilibrium, boudet2025non}.
This behavior reflects the non-Markovian nature of the reduced dynamics $x(t)$.

In contrast, the bidimensional embedding Eq.~(\ref{Eq:Langevin}) provides a Markovian description which allows generalized FRR to be applied.
This is conditioned on the identification of the correct conjugate variable $Y(x,\eta)$ to the perturbation applied \cite{prost2009generalized, dinis2012fluctuation, engbring2023nonlinear}.
The response function of a given observable is then related to the correlation of $Y$ with the same observable \cite{prost2009generalized}.
The conjugate variable is defined from the Markovian-embedding steady-state $p(x,\eta)$ and, for a small perturbation, can be linearized
\begin{equation}
Y(x,\eta) = \frac{1}{p(x,\eta)} \frac{\partial p(x,\eta)}{\partial \kappa}.
\label{Eq:ConjVar}
\end{equation}
We further define the dimensionless conjugate variable $\widetilde Y = \kappa_{\rm f} Y$, leading to the following generalized FRR
\begin{equation}
    \mathscr{R}(t) = \frac{\Delta \kappa}{\kappa_{\rm f}} \langle \widetilde Y(0) x^2(t)\rangle_{\rm f}
    \label{Eq:GFRR}
\end{equation}
as we demonstrate is Appendix \ref{App:GFRR}.
This result is equivalent to Eq.(5) in Ref.~\cite{prost2009generalized} using $x^2$ as an observable, instead of the conjugate variable itself.
In the general case, the conjugate variable can be evaluated directly from the measured histograms \cite{engbring2023nonlinear}.
In our case, the bidimensional state $p(x,\eta)$ can be expressed as a function of $x$ and $\eta$ \cite{crisanti2012nonequilibrium} which leads to an explicit expression of $\widetilde Y$.
This allows to express the r.h.s. of  Eq.~(\ref{Eq:GFRR}) as a function of three time-dependent correlation functions $\langle x^2(0) x^2(t)\rangle_{\rm f}$,  $ \langle x(0)\eta(0)x^2(t) \rangle_{\rm f}$ and  $\langle \eta^2(0) x^2(t) \rangle_{\rm f}$, each evaluated in the final NESS.
The result is a simple yet lengthy equation, which is presented in the Appendix \ref{App:Explict}.
Each correlation function can be measured experimentally and is shown in Fig.~\ref{fig:ConjVar} (Appendix \ref{App:Explict}), in good agreement with the result of numerical simulations.
This explicit form of r.h.s. of Eq.~(\ref{Eq:GFRR}) allows to test experimentally the generalized FRR with high resolution.

In Fig.~\ref{fig:FRR}(c) we show the response function $\mathscr{R}(t)$ together with the scaled correlation function $\frac{\Delta \kappa}{\kappa_{\rm f}} \langle \widetilde Y(0) x^2(t)\rangle_{\rm f}$ of the conjugate variable $\widetilde Y$ with the observable $x^2(t)$.
The agreement between both quantities shows that the generalized FRR [Eq.~(\ref{Eq:GFRR})] holds when the appropriate conjugate variable is evaluated in the Markovian embedding.
This bidimensional description is sufficient to recover a simple relation between the response of the observable $x^2$ of the system to a mechanical $\kappa$-change and its correlation with the conjugate variable $\widetilde Y$.
The practical implication of this result is the ability to predict the complex time-evolution of the system after a stiffness-perturbation in the presence of colored noise.
This prediction relies on the evaluation of the appropriate correlation function, implying a nontrivial conjugate variable.\\

\textit{Conclusion}\---\
In this work, we experimentally demonstrate how generalized fluctuation–response relation apply to nonequilibrium system when the appropriate Markovian embedding is resolved.
Our experimental setup, which involves an optically trapped microparticle driven by colored noise, enables us to fully resolve the required Markovian embedding.
We focus on the system's response to a mechanical change in stiffness, which modifies the system’s state beyond its mean position.
This type of perturbation applied to a nonequilibrium state represents a first step toward describing dynamical changes in microbiological processes.
It demonstrates that resolving the Markovian embedding enables to predict the nonequilibrium response from steady-state fluctuations, even for a complex systems subjected to a changes in mechanical stiffness.
More generally, this validation supports the use of Markovian embeddings as an operational basis to apply FRRs to nonequilibrium systems in which the additional degrees of freedom are not directly accessible but can be inferred from measured trajectories \cite{dinis2012fluctuation}.

\acknowledgements
Y.R. and R.G. acknowledge support from the European Research Council (ERC) under the European Union’s Horizon 2020 research and innovation program (Grant Agreement No. 101002392).
Y.R. acknowledges support from the Israel Science Foundation (grants No. 385/21, 2013/25).
A.T. acknowledges financial support from QuantTEdu-France (Project No. ANR-22-CMAS-0001).
This work is also part of the Interdisciplinary Thematic Institute QMat of the University of Strasbourg, CNRS, and Inserm, supported by the following programs: IdEx Unistra (Grant No. ANR- 10-IDEX-0002), SFRI STRATUS Project (Project No. ANR-20-SFRI0012), under the framework of the French Investments for the Future Program.

\subsection*{Data availability}
All experimental data used in the work as well as the codes necessary to perform the present analysis will be made available online.


%

\appendix

\section{Experimental Setup}
\label{App:Exp}

\begin{figure}[t!]
	\centerline{\includegraphics[width=0.85\linewidth]{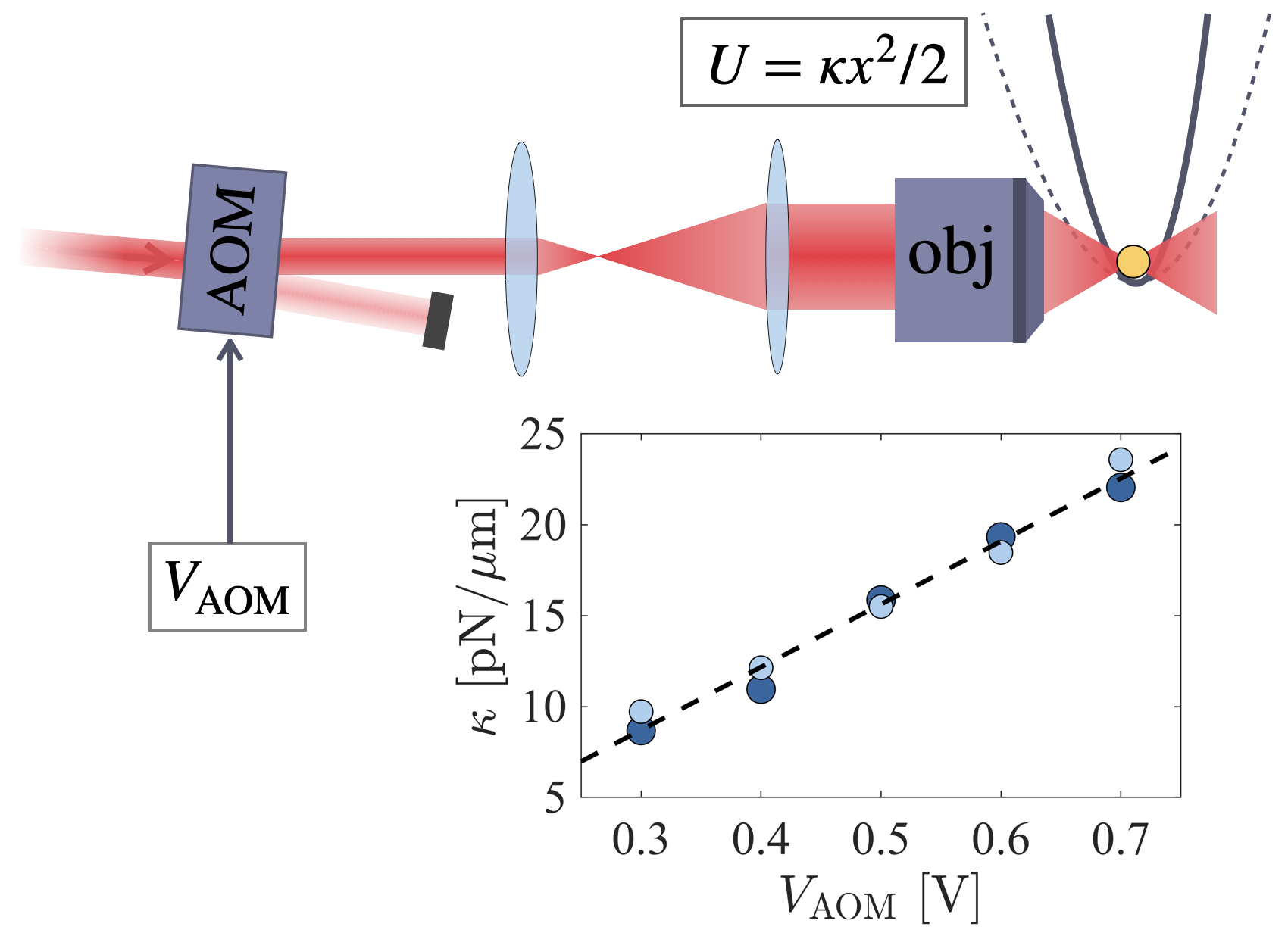}}
	\caption{Schematic view of the calibration of the optical potential. The trapping laser is sent through an acousto-optic modulator (AOM). The AOM is fed with a tension $V_{\rm AOM}$  which controls the intensity of the first order diffracted beam, that is sent into an high NA objective to form the optical potential. Modifying $V_{\rm AOM}$ leads to a change in the trap stiffness $\kappa$.
    The linear relation between $\kappa$ and $V_{\rm AOM}$ is calibrated (lower right inset), allowing to control precisely the stiffness of the potential.}
	\label{fig:calib}
\end{figure}

Our platform (detailed in Ref.~\cite{goerlich2022harvesting}) consists of optically trapping, in a harmonic potential, a single dielectric bead ($3~\rm{\mu m}$ polystyrene sphere) in a fluidic cell filled with deionized water at room temperature $T = 296$ K.
The harmonic potential is induced by focusing inside the cell a linearly polarized Gaussian beam ($785$ nm, CW $110$ mW laser diode, Coherent OBIS) through a high numerical aperture (NA) objective (Nikon Plan Apo VC, $60\times$, NA$=1.20$ water immersion). An additional force in the form of radiation pressure is applied to the sphere using a time-dependent fraction of the light-beam emitted by an additional high-power laser ($800$ nm, CW $5$ W Ti:Sa laser, Spectra Physics 3900S).
It is sent on the micrsophere as a thin beam, strongly underfilling the aperture of a lower-NA objective (Nikon Plan Fluor Extra Large Working Distance, $60\times$, NA$=0.7$) in order to prevent additional gradient forces.
The intensity of both the trapping laser and the radiation pressure beam are controlled by two independent acousto-optic modulators (Gooch and Housego 3200) using a digital-to-analogue card (NI PXIe 6361) and a \textsc{python} code.
The AOM controlling the trapping laser is calibrated directly from the corner frequency of the power spectral density of the recorded bead dynamics (see Fig.~\ref{fig:calib}), allowing for a control of the stiffness $\kappa$.
This allows to control independently the stiffness of the optical potential, which is linearly related to the voltage sent to the first AOM, and the amplitude of the time-dependent external force with the second AOM \cite{goerlich2022harvesting}.

The instantaneous position $x(t)$ of the sphere along the optical axis is measured by recording the light scattered off the sphere of a low-power $639$ nm laser (CW $30$ mW laser diode, Thorlabs HL6323MG), sent on the bead via the second, low-NA objective. The scattered light is collected by the trappeing beam objective and recorded by a photodiode ($100$ MHz, Thorlabs Det10A).
The recorded signal (in V/s) is amplified using a low noise amplifier (SR560, Stanford Research) and then acquired by an analog-to-digital card (NI PCI-6251). The signal is filtered through a $0.3$ Hz high-pass filter at 6 dB/oct to remove the DC component and through a $100$ kHz low-pas filter at 6 dB/oct to prevent aliasing. The scattered intensity varies linearly with the position of the trapped bead $x(t)$ for small enough displacements, and we make sure to work in the linear response regime of the photodiode so that the recorded signal is linear with the intensity, resulting in a voltage trace $v(t)$ well linear with the position $x(t)$ of the microsphere in the trap.

\section{Explicit form of the generalized FRR}
\label{App:Explict}

The conjugate variable $\widetilde Y(x,\eta)$ given by Eq.~(\ref{Eq:ConjVar}) depends on the bidimensional state $p(x,\eta)$ which is known in our case \cite{crisanti2012nonequilibrium, dinis2012fluctuation}. This allows for the direct expression
\begin{equation}
    \widetilde Y(x,\eta) = a_0 + a_{11} x^2 + a_{12} x\eta + a_{22}\eta^2
\end{equation}
with coefficients [using $M \equiv D \gamma^2 (\tau_{\rm f} + \tau_c)^2 +  \sigma_\eta^2 \tau_{\rm f}^2 \tau_c$]:
\begin{align*}
a_0 &= \frac{D \gamma^{2} (\tau_{\rm f} + \tau_c)^{3}
      +  \sigma_\eta^2 \tau_{\rm f} \tau_c^{2} (\tau_{\rm f} + 3\tau_c)}
      {2 (\tau_{\rm f} + \tau_c) M},
\\[1ex]
a_{11} &= - \frac{\gamma^{2} (\tau_{\rm f} + \tau_c) 
          \bigl(D \gamma^{2} [\tau_{\rm f} + \tau_c]^{3}
          +  \sigma_\eta^2 \tau_{\rm f} \tau_c^{2} [\tau_{\rm f} + 3\tau_c]\bigr)}
          {2 \tau_{\rm f} M^{2}},
\\[1ex]
a_{12} &= \frac{\gamma \tau_{\rm f} \tau_c 
          \bigl(D \gamma^{2} [\tau_{\rm f} + \tau_c]^{2}
          +  \sigma_\eta^2 \tau_{\rm f} \tau_c [\tau_{\rm f} + 2\tau_c]\bigr)}
          {M^{2}},
\\[1ex]
a_{22} &= - \frac{D \gamma^{2} \tau_{\rm f} \tau_c^{2} 
          (\tau_{\rm f}^{2} - \tau_c^{2})
          +  \sigma_\eta^2 \tau_{\rm f}^{3} \tau_c^{3}}
          {2 M^{2}} .
\end{align*}

Using this form of the conjugate variable in cross correlation $\langle\widetilde Y(0) x^2(t)\rangle$ leads to the generalized FRR
\begin{equation}
\begin{aligned}
        \mathscr{R}(t) &= \frac{\Delta \kappa}{\kappa_{\rm f}} \bigg[ a_0\langle x^2(t) \rangle_{\rm f} + a_{11} \langle x^2(0) x^2(t)\rangle_{\rm f} \\
         &+ a_{12} \langle x(0)\eta(0)x^2(t) \rangle_{\rm f} + a_{22} \langle \eta^2(0) x^2(t) \rangle_{\rm f} \bigg].
    \end{aligned}
\label{Eq:GFRR_explicit}
\end{equation}

\begin{figure}[t!]
	\centerline{\includegraphics[width=1\linewidth]{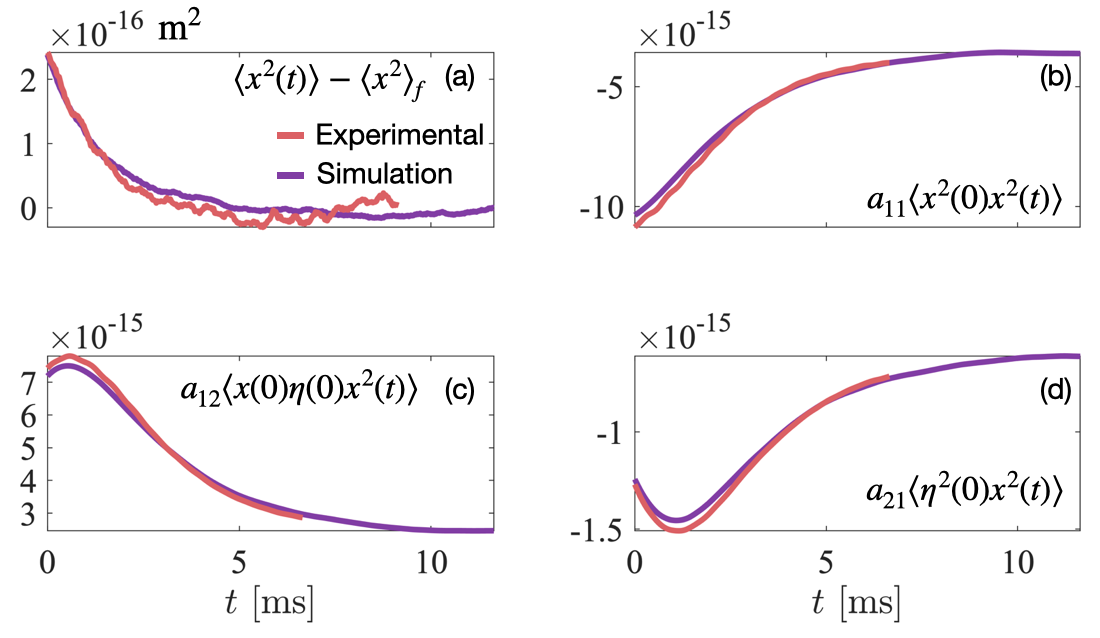}}
	\caption{Experimental results (red lines) and results of numerical simulation with the experimental parameters (purple lines) in $\rm{m^2}$ for 
    (a) the response function $\mathscr{R}(t)$,
    (b) the scaled autocorrelation $\langle x^2(0) x^2(t) \rangle_{\rm f}$
    (c) the scaled cross-correlation $\langle x(0) \eta(0) x^2(t) \rangle_{\rm f}$
    and (d) the scaled cross-correlation $\langle \eta^2(0) x^2(t) \rangle_{\rm f}$, the latter three being evaluated in the final steady-state.}
	\label{fig:ConjVar}
\end{figure}

The response function $\mathscr{R}(t)$ as well as the three cross-correlations on the r.h.s. of Eq~(\ref{Eq:GFRR_explicit}) can be measured experimentally (Fig.~\ref{fig:ConjVar}).
The correlation functions are evaluated as mixed time and ensemble average over the ensemble of trajectories in the final steady state.
The experimental results are complemented by numerical simulations of the bidimensional process, Eq.~(\ref{Eq:Langevin}), using the experimentally measured parameters.
The explicit form of the FRR [Eq.~(\ref{Eq:GFRR_explicit})] composed of these measured cross-correlations is the result shown in Fig.~\ref{fig:FRR}(c) in the main text.

\section{Response function in the presence of colored noise}
\label{App:Var}

As shown in Ref.~\cite{Goerlich2025FT}, the time evolution of the second moment $\sigma^2(t) = \langle x^2(t) \rangle$ after a change of stiffness from $\kappa_{\rm i}$ to $\kappa_{\rm f}$ reads
\begin{equation}
\label{eq:NeqVar_tau}
\begin{gathered}[b]
\sigma^2(t) =
\left( D \tau_{\rm i} + \frac{\sigma_\eta^2 \tau_{\rm i}^2 \tau_c}
{\gamma^2 (\tau_{\rm i} + \tau_c)} \right)
e^{-2 t/\tau_{\rm f}}
+ D \tau_{\rm f}\left(1 - e^{-2 t/\tau_{\rm f}}\right)
\\
+ \frac{\sigma_\eta^2}{\gamma^2}
\left(
\frac{\tau_{\rm f}^2 \tau_c}{\tau_{\rm f} + \tau_c}
+ \frac{2 \tau_{\rm f}^2 \tau_c}{\tau_{\rm f} - \tau_c} e^{-2 t/\tau_{\rm f}}
- \frac{2 \tau_{\rm f}^2 \tau_c^2}{\tau_{\rm f}^2 - \tau_c^2} 
e^{-t\left(\frac{1}{\tau_{\rm f}}+\frac{1}{\tau_c}\right)}
\right)
\\
+ \frac{2\sigma_\eta^2 \tau_{\rm i} \tau_{\rm f} \tau_c}
{\gamma^2(\tau_{\rm i} + \tau_c)(\tau_{\rm f} - \tau_c)}
\left(
e^{-t\left(\frac{1}{\tau_{\rm f}}+\frac{1}{\tau_c}\right)}
- e^{-2 t/\tau_{\rm f}}
\right),
\end{gathered}
\end{equation}
where $\tau_{\rm f} \equiv \gamma / \kappa_{\rm f}$.
This enters the response function $\mathscr{R}(t) = \sigma^2(t) - \sigma_{\rm f}^2$.
In the absence of colored noise $\sigma_{\eta}$, one retrieves the standard result $\sigma^2(t) = (\sigma_{\rm i} - \sigma_{\rm f})e^{-2t/\tau_{\rm f}} + \sigma_{\rm f}$ valid for a transient between two equilibrium states.

\section{Generalized Fluctuation-Response Relation for Markovian dynamics}
\label{App:GFRR}

In what follows we consider a simple generalization of a Fluctuation-dissipation relation within a class A approach according to the classification of Ref.\cite{caprini2021fluctuation}. This is essentially the
same derivation as is discussed in the Supplemental Material to Ref.~\cite{sokolov2023linear}, but uses a different notation, which is compatible with the notation in the rest of the present work. The FRR which we derive in this way is related to the FDT of Prost et al \cite{prost2009generalized} obtained via a different approach. The only necessary assumptions for our
derivation are the Markovian nature of the system's evolution, and the existence of the stationary state for any magnitude of perturbation in the admissible range. 

Parallel to the standard discussion for the equilibrium, we prepare a system in a stationary state under the action of the constant perturbation $\phi$, which is switched off at time $t=0$, after which the system undergoes autonomous evolution. The stationary state under constant perturbation $F$ is characterized by the probability density function (PDF) $p(\mathbf{x}|\phi)$, where $\mathbf{x}$ is a vector of internal state variables of the system. The number of these variables must be taken to be sufficient for Markovian description of its time evolution. 
In our case, the perturbation is a stiffness modification $\phi = (\kappa_{\rm f}-\kappa_{\rm i})$.

The PDF of $\mathbf{x}$ at time $t$ after switching of the perturbation $\phi$ is given by
\[
 p(\mathbf{x};t) = \int p(\mathbf{x},t|\mathbf{x}';0) p(\mathbf{x}'|\phi) d \mathbf{x}',
\]
where $p(\mathbf{x},t|\mathbf{x}';t')$ denotes the transition probability density from the state $\mathbf{x}'$ at time $t'<t$ to the state $\mathbf{x}$ at time $t$. 
Now we consider a single-time observable $\mathcal{O}(t)\equiv\mathcal{O}[x(t)]$. 
Its mean value at time $t$ is given by
\[
\langle \mathcal{O}(t)| \phi \rangle = \int \! \!\! \! \int  \mathcal{O}(\mathbf{x}) p(\mathbf{x},t|\mathbf{x}',0) p(\mathbf{x}'|\phi) d\mathbf{x} d\mathbf{x}'.
\]
This mean is then compared to a reference state, in which the system is prepared without perturbation (i.e. with $\phi = 0$):
\[
\langle \mathcal{O}(t)| 0 \rangle = \int \! \!\! \! \int  \mathcal{O}(\mathbf{x}) p(\mathbf{x},t|\mathbf{x}',0) p(\mathbf{x}'|0) d\mathbf{x} d\mathbf{x}'.
\]
The response of the system to the perturbation is thus given by 
\begin{eqnarray*}
&& \langle \mathcal{O}(t)| \phi \rangle - \langle \mathcal{O}(t)| 0 \rangle = \int  \! \!\! \!  \int d\mathbf{x} d\mathbf{x}' \mathcal{O}(\mathbf{x}) p(\mathbf{x},t|\mathbf{x}',0)\\
&& \qquad \times  [p(\mathbf{x}'|\phi) - p(\mathbf{x}'|0)].
\end{eqnarray*}
Introducing 
\[
 \mathcal{Y}(\mathbf{x}, \phi) = \frac{p(\mathbf{x}|\phi) - p(\mathbf{x}|0)}{p(\mathbf{x}|0)}
\]
we thus may write 
\begin{eqnarray*}
 && \langle \mathcal{O}(t)| \phi \rangle - \langle \mathcal{O}(t)| 0 \rangle =  \\
 && \qquad \int  \! \!\! \!  \int d\mathbf{x} d\mathbf{x}'  \mathcal{O}(\mathbf{x}) \mathcal{Y}(\mathbf{x}',\phi) p(\mathbf{x},t|\mathbf{x}',0) p(\mathbf{x}'|0) = \\
 && \qquad \int  \! \!\! \!  \int d\mathbf{x} d\mathbf{x}'  \mathcal{O}(\mathbf{x}) \mathcal{Y}(\mathbf{x}',\phi) p(\mathbf{x},t;\mathbf{x}',0) \\
\end{eqnarray*}
where $p(\mathbf{x},t;\mathbf{x}',0)$ is the joint PDF of internal variables at times $t$ and $0$. Thus,
\begin{equation}
\langle \mathcal{O}(t)| \phi \rangle - \langle \mathcal{O}(t)| 0 \rangle =  \left\langle  \mathcal{O}(t) \mathcal{Y}(\phi,0) \right\rangle_0
 \label{eq:FDTL}
\end{equation}
By writing $ \mathcal{Y}(\phi, 0)$ we stress that the variable  $\mathcal{Y}$ (the \textit{exact conjugate} variable to the perturbation) is calculated or measured at $t=0$. The subscript 0 in the last 
mean indicates that this one is calculated using the joint PDF in the absence of the perturbation. Note that $\mathcal{Y}$ itself is still dependent on the perturbation strength. 
Eq.~(\ref{eq:FDTL}) is the nonlinear fluctuation-dissipation theorem of Ref.~\cite{engbring2023nonlinear}.

Now, let us assume that $p(\mathbf{x}|\phi)$ is differentiable w.r.t. $\phi$, and approximate $ \mathcal{Y}(\mathbf{x},\phi)$ in the linear order:
\begin{eqnarray*}
  \mathcal{Y}(\mathbf{x}, \phi) &\simeq & \frac{1}{p(\mathbf{x}|0)} \left. \frac{d}{d\phi} p(\mathbf{x}|\phi)\right|_{\phi=0} \phi \\
  & =& \phi \left. \nabla_\phi \ln p(\mathbf{x},\phi) \right|_{\phi=0} \equiv \phi Y(\mathbf{x}),
\end{eqnarray*}
where the variable $Y(\mathbf{x})$ is now independent from the perturbation's magnitude. We then get:
\begin{equation}
\langle \mathcal{O}(t)| \phi \rangle - \langle \mathcal{O}(t)| 0 \rangle =  \phi \left\langle  \mathcal{O}(t) Y(0) \right\rangle_0 .
 \label{eq:FDTStan}
\end{equation}
Eq.~(\ref{eq:FDTStan}) is the customary FRR in a stationary state (equilibrium or not). Taking $\mathcal{O}(\mathbf{x}) = Y(\mathbf{x})$ we get the FDT of Prost et al \cite{prost2009generalized}
(up to a different sign convention). This FRR is pertinent to the linear response case.

In equilibrium, this relation 
\[
 \langle Y(t)| \phi \rangle - \langle Y(t)| 0 \rangle =  \phi \left\langle Y(t) Y(0) \right\rangle_0
\]
it is the standard FDT. 
In our case, focusing on the observable $\mathcal{O}(t) = x^2$ leads to
\begin{equation}
    \langle x^2(t)|\phi \rangle - \langle x^2(t) |0 \rangle = \phi \langle x^2(t) Y(0) \rangle
\end{equation}
with the l.h.s. being the response function of $x^2$ to the perturbation $\phi$.

\end{document}